\documentclass[a4paper, 11pt]{article}
\usepackage{graphicx}
\usepackage[usenames]{color}
\usepackage[flushleft]{threeparttable}
\DeclareGraphicsExtensions{.pdf,.png,.jpg,.mps,.eps,.ps}
\usepackage{amsmath,amssymb,gensymb}
\usepackage{natbib} 
\setcitestyle{authoryear,open={(},close={)}}
\usepackage{color}

\usepackage{lscape}

\usepackage{psfrag}
\usepackage[colorlinks,citecolor=blue,urlcolor=blue,bookmarks=false,hypertexnames=true,pagebackref=true]{hyperref}

\newcommand\nicer{{\it NICER}}
\newcommand\nustar{{\it NuSTAR}}
\newcommand\swift{{\it SWIFT}}

\newcommand\chandra{{\it Chandra}}

\newcommand\rxte{{\it RXTE}}
\newcommand\xmm{{\it XMM-Newton}}
\newcommand\inte{{\it INTEGRAL}}
\newcommand\maxi{{\it MAXI}}

\newcommand\kev{{\rm~keV}}

\newcommand\kms{\ifmmode {\rm~km\ s}^{-1} \else ~km s$^{-1}$\fi}
\newcommand\Hunit{\ifmmode {\rm~km\ s}^{-1}\ {\rm Mpc}^{-1}
        \else ~km s$^{-1}$ Mpc$^{-1}$\fi}
\newcommand\ctssec{\ifmmode {\rm~count\ s}^{-1} \else ~count s$^{-1}$\fi}
\newcommand\ergsec{\ifmmode {\rm~erg\ s}^{-1} \else
        ~erg s$^{-1}$\fi}
\newcommand\funit{\ifmmode {\rm~erg\ s}^{-1}\;{\rm cm}^{-2} \else
        ~ergs s$^{-1}$ cm$^{-2}$\fi}
\newcommand\phflux{\ifmmode {\rm~photon\ s}^{-1}\;{\rm cm}^{-2}
        \else   ~photon s$^{-1}$ cm$^{-2}$\fi}
\newcommand\efluxA{\ifmmode {\rm~erg\ s}^{-1}\;{\rm cm}^{-2}\;{\rm
        \AA}^{-1} \else ~erg s$^{-1}$ cm$^{-2}$ \AA$^{-1}$\fi}
\newcommand\efluxHz{\ifmmode {\rm~erg\ s}^{-1}\;{\rm cm}^{-2}\;{\rm
        Hz}^{-1} \else ~erg s$^{-1}$ cm$^{-2}$ Hz$^{-1}$\fi}
\newcommand\cc{\ifmmode {\rm~cm}^{-3} \else cm$^{-3}$\fi}
\newcommand\FWHM{\ifmmode {\rm~FWHM} \else ${\rm~FWHM}$\fi}
\newcommand\Msun{\ifmmode M_{\odot} \else $M_{\odot}$\fi}
\newcommand\Lsun{\ifmmode L_{\odot} \else $L_{\odot}$\fi}

\newcommand\hbeta{\ifmmode {\rm H}\beta \else H$\beta$\fi}
\newcommand\Kalpha{\ifmmode {\rm K}\alpha \else K$\alpha$\fi}
\newcommand\nh{\ifmmode N_{\rm H} \else N$_{\rm H}$\fi}

\title{\nustar{} view of the X-ray transients Swift J174805.3-244637 and IGR~J17511-3057}
\author{Aditya S. Mondal$^{1}$\thanks{E-mail: adityas.mondal@visva-bharati.ac.in}, Mahasweta Bhattacharya$^{1}$, Mayukh Pahari$^{2}$,\\
  Biplab Raychaudhuri$^{1}$, Rohit Ghosh$^{1}$, Gulab C. Dewangan$^{3}$   \\
  {\small
$^{1}$ Department of physics, Visva-Bharati, Santiniketan, West Bengal, 731235, India }\\
{\small $^{2}$ Department of Physics, Indian Institute of Technology Hyderabad},\\{\small Hyderabad, Kandi, 502285 Sangareddy, India}\\
{\small $^{3}$ Inter-University Centre for  Astronomy \& Astrophysics (IUCAA), Pune, 411007, India} \\
}

\date{\today}
\begin{document}
\pagestyle{empty}
\maketitle
\begin{abstract}
We report on the \nustar{} observations of the neutron star low-mass X-ray binary Swift J174805.3-244637 (hereafter Swift~J17480) and the accreting millisecond X-ray pulsar IGR~J17511-3057 performed on March 4, 2023, and April 8, 2015, respectively. We describe the continuum emission of Swift~J17480 with a combination of two soft thermal components and an additional hard X-ray emission described by a power-law. We suggest that the spectral properties of Swift~J17480 are consistent with a soft spectral state. The source IGR~J17511-3057 exhibits a hard spectrum characterized by a Comptonized emission from the corona. The X-ray spectrum of both sources shows evidence of disc reflection. For the first time, we employ the self-consistent reflection models ({\tt relxill} and {\tt relxillNS}) to fit the reflection features in the \nustar{} spectrum. From the best-fit spectral model, we find an inner disc radius ($R_{in}$) is precisely constrained to $(1.99-2.68)\:R_{ISCO}$ and inclination to $30\pm 1\degree$ for Swift~J17480. We determine an inner disc radius of $\lesssim 1.3\;R_{ISCO}$  and inclination of $44\pm 3\degree$ for IGR~J17511-3057. A low inclination angle of the system is required for both sources. For the source IGR~J17511-3057, spinning at $4.1$ ms, the value of co-rotation radius ($R_{co}$) is estimated to be $\sim 42$ km ($3.6\:R_{ISCO})$, consistent with the position of inner disc radius as $R_{in}\lesssim R_{co}$. We further place an upper limit on the magnetic field strength of the sources, considering the disc is truncated at the magnetospheric radius.

 \end{abstract}


\section{Introduction}
Neutron star (NS) Low-mass X-ray binaries (LMXBs) consist of an NS and a companion star that is less massive ($\lesssim 1 \Msun$) than the compact primary. The companion typically overflows its Roche-lobe, feeding gas to an accretion disk surrounding the NS \citep{2006csxs.book..623T}. The presence of the NS surface interrupts the accretion flow, forming a boundary or spreading layer between the disk and the NS surface, extending to high latitudes at higher accretion rates \citep{2001ApJ...547..355P}. NS LMXBs are classified into persistent and transient sources based on long-term X-ray variability. Persistent NS LMXBs accreate matter continuously and may have an X-ray luminosity of $L_{X}\gtrsim 10^{36} \ergsec{}$ (\citealt{2019ApJ...873...99L, 2017ApJ...836..140L}). Meanwhile, transient NS LMXBs undergo occasional outbursts during which matter from the less massive companion is rapidly accreted onto the NS. Typical outbursts reach a luminosity of $L \simeq 10^{36-38}$ erg s$^{-1}$ and last for weeks to months. Outbursts are usually separated by long (years to decades) periods of quiescence with a much lower X-ray luminosity of $L\simeq 10^{31-34}$ erg s$^{-1}$ \citep{2010A&A...524A..69D, 2020arXiv201009005D}. The mass-accretion rate onto the NS is strongly reduced during the quiescence. This transient behaviour is thought to be due to thermal-viscous instabilities in the accretion disc \citep{2001NewAR..45..449L}.\\

An accreting pulsar is an NS accreting mass from a low-mass donor/companion star. The sources are known as accreting millisecond pulsars (AMSPs) when they are rotating at millisecond periods, and coherent pulsations occur in the persistent X-ray emission \citep{2018ASSL..457..149C, 2022ASSL..465...87D}. Their pulsations are thought to be powered by accretion, when matter from the accretion disc is channeled by the magnetic field lines onto the magnetic pole, forming a hotspot visible in X-rays. As a binary system evolves through accretion phases onto NS, it gains angular momentum from the accreted material, sufficient to spin up the NS to a rotation period equilibrium in the millisecond range. Studying the frequency variations of the coherent signal emitted by an accreting pulsar is one of the foremost techniques for studying the dynamic interaction between the rotating NS and the accreted matter. Only 27 AMSPs have been discovered since 1998. SAX~J1808.4-3658 is the first discovered AMSP \citep{1998ApJ...507L..63W} with a spin frequency of $401$ Hz, the recent ones being MAXI~J1816-195 \citep{2022ATel15425....1B},  MAXI~J1957+032 \citep{2022ATel15444....1N}, and SRGA~J144459.2-604207 \citep{2024ATel16464....1M}. Most AMSPs are transient X-ray sources with recurrence times between two and more than ten years, and their outbursts usually last from a week to a few months \citep{Patruno_2020}. The spin frequencies of AMSPs lie in the range of $180-600$ Hz. Almost all AMSPs have short, i.e., $\sim$hours or minutes (between 40 mins and 5 h), orbital periods and are therefore characterized by compact orbits. Most of these systems are bursters, as they have displayed a type-I thermonuclear X-ray burst at least once (\citealt{2021ASSL..461..209G, 2010MNRAS.407.2575P, 2019MNRAS.483..767D, 2022ApJ...935L..32B, 2024ApJ...968L...7N}). The bursts from AMSPs exhibit burst oscillations at the typical spin frequency \citep{Patruno_2020}. The AMSPs also exhibit quasi-periodic oscillations \citep{2003Natur.424...42C}. The magnetic field of the NS, as inferred from accretion models, is relatively weak, in the range of $\sim 10^{8}-10^{9}$ G \citep{1999ApJ...521..332P}. \\

The AMSPs exhibit many characteristics similar to other low-mass X-ray binaries (LMXBs). Their broadband spectra show soft thermal and hard Comptonized components similar to atoll sources in the low/hard state \citep{2006AdSpR..38.2697P}. The broadband continuum of AMSPs has been invariably found to be dominated by the Comptonization spectrum from a hot corona with electron temperatures usually of tens of keV \citep{2022ASSL..465...87D}. However, two AMSPs, SAX~J1748.9-2021 and SAX~J1808.4-3658, have been observed to transition into the soft states \citep{2016MNRAS.457.2988P, 2019MNRAS.483..767D}. Two soft thermal components with relatively low temperatures are often detected in the spectra of AMSPs besides an energetically dominating Comptonized component. The cooler one is attributed to the accretion disc emission, while the hotter one is interpreted as the thermal emission of the NS surface. An additional spectral component arises when the Comptonization spectrum emitted by the corona illuminates the disc and is reprocessed by it, known as the reflection spectrum (\citealt{1989MNRAS.238..729F, 2024Ap&SS.369...16L} for review). Disk reflection components are mainly characterized by a broad emission line in the $\simeq 6.4–6.97$ keV band due to iron plus a Compton back-scattering hump at $\simeq 20–40$ keV \citep{1989MNRAS.238..729F}. Disc reflection spectra may provide a powerful probe of the accretion geometry, such as the inner radial extent of the accretion disk, the ionization of the disc's plasma, and the system's inclination. Reflection features have been observed in some AMSPs for which data with high-to-moderate energy resolution were available \citep{2013MNRAS.429.3411P, 2017MNRAS.471..463S, 2019MNRAS.483..767D, 2022MNRAS.515.3838M} but not all \citep{2005A&A...436..647F, 2018A&A...610L...2S, 2018A&A...616L..17S}. \\

\subsection{Swift J174805.3-244637 (Swift~J17480)}
Terzan 5 is a dense and massive globular cluster close to the center of our Milky Way Galaxy ($d = 5.9 \pm 0.5$ kpc; \citealt{2007AJ....133.1287V}). It harbors a large number of faint X-ray point sources, including a dozen likely quiescent neutron star LMXBs \citep{2006ApJ...651.1098H}. X-ray outburst activities have been observed from this cluster on many occasions since 1980 (\citealt{1981ApJ...247L..23M, 2012MNRAS.422..581D}). However, it was difficult to detect with certainty the object(s) which caused these outbursts because of the high source density and lack of sub-arcsecond spatial resolution observations. Previously, two confirmed transient NS LMXBs were identified in the cluster. Those were EXO~1745–248 and the 11-Hz X-ray pulsar IGR~J17480–2446. The source EXO~1745–248 shows its X-ray activity at least in 2000, 2011, and 2015 (\citealt{2003ApJ...590..809H, 2012MNRAS.426..927A, 2015ATel.7240....1A}), whereas, the source IGR~J17480–2446 was responsible for the 2010 outburst of Terzan 5 and no X-ray outburst activity has been observed from this source in later time (\citealt{2010ATel.2919....1B, 2010ATel.2940....1F, 2010ATel.2933....1H, 2010ATel.2974....1P, 2010ATel.2929....1S}). \\

\begin{figure*}
\centering
\includegraphics[width=0.75\textwidth, angle=0]{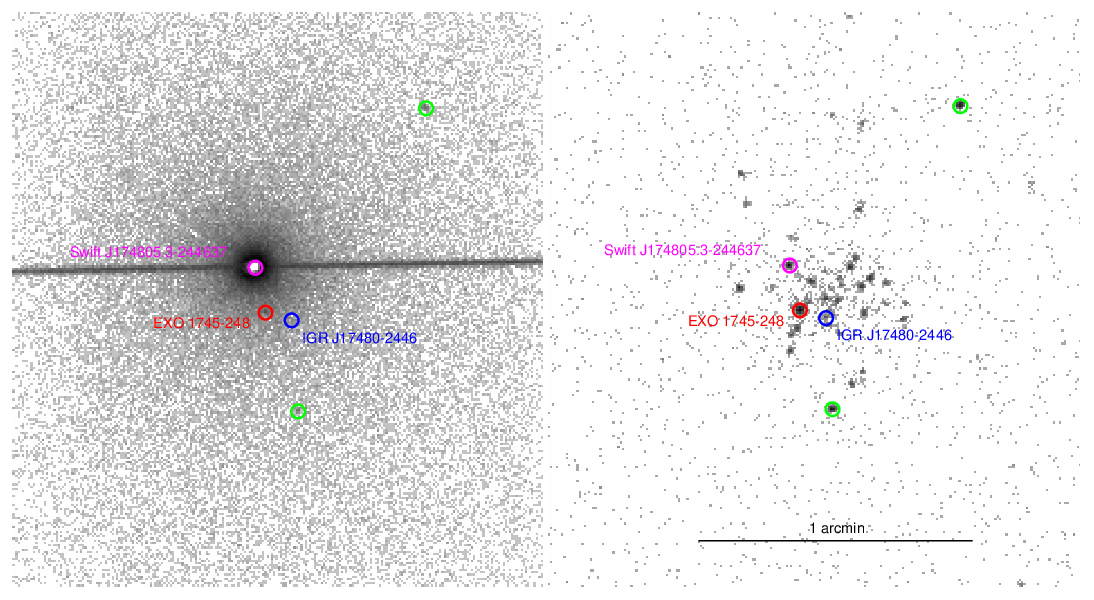}
\caption{ Left panel shows the position of the known X-ray transients in the Terzan 5 globular cluster using the 10.6 ks \chandra{} observation taken on March 18, 2023, during it's current outburst. Right panel shows the same but uses a 36.26 ks \chandra{} observation from July 15, 2009. The three known X-ray transients and two other visible \chandra{} X-ray sources have been marked and labeled. In the left image,  the same region file has been opened and shifted by a fraction of an arcsecond to align the three other visible X-ray sources. This leads to a confident identification of Swift~J174805.3-24467 as the transient active in 2023 \citep{2023ATel15953....1H}. }
\label{Fig0}
\end{figure*}

In 2012, a third transient NS LMXB was discovered in Terzan 5: Swift J174805.3–244637, hereafter referred to as Swift~J17480 (also known as Terzan 5 X-3; \citealt{2014ApJ...780..127B}). During a \swift{} observation taken on July 17, 2012, \citet{2012ATel.4264....1A} detected a type-I X-ray burst in a $\sim 950$ seconds long observation, which allowed them to identify Swift~J17480 as the 3rd conclusively identified NS LMXB in Terzan 5. The source remained active for $7-8$ weeks and then transitioned to quiescence \citep{2014ApJ...780..127B}. Recently, a second (new) outburst from this transient was observed on February 27, 2023, by \maxi{}/GSC (\citealt{2023ATel15917....1N, 2023ATel15919....1K}). The outburst was also reported by \inte{} \citep{2023ATel15921....1F}. The \nicer{} observation performed on February 28, 2023, also reported on the detection of the source Swift~J17480 \citep{2023ATel15922....1S}. Later, using \chandra{} observation performed on March 19, 2023, \citet{2023ATel15953....1H} identified the current active transient as Swift~J17480 at high confidence. Also, the timing of this outburst ($\sim 11$ years after Swift~J17480's 2012 outburst) matches the prediction of an 8-10 year recurrence time suggested by \citet{2014ApJ...780..127B}. The cluster core with the locations of the three transient LMXBs is shown in Figure~\ref{Fig0} using a 10.6 ks \chandra{} observation on March 18, 2023 \citep{2023ATel15953....1H}. \\

During the 2012 outburst, the source was bright for approximately 20 days, reaching a maximum luminosity of $\sim 7\times 10^{37}$ erg~s$^{-1}$ and an average luminosity of $\sim 3\times 10^{37}$ erg~s$^{-1}$ in the $0.5–10$ keV band in this time interval, assuming a distance of $5.9$ kpc \citep{2014ApJ...780..127B}. \swift{} observation of the 2012 outburst detected a thermonuclear type-I X-ray burst, the spectrum of which is well-fitted with an absorbed blackbody model. Time-resolved spectral analysis revealed that the blackbody temperature decreases from $\sim 2.3$ keV at the peak to $1.2$ keV as the burst evolves from peak to persistent emission \citep{2012ATel.4264....1A}. \citet{2014ApJ...780..127B} also identified a Type I X-ray burst in \swift{}/XRT data with a long (16 s) decay time, indicative of hydrogen burning on the surface of the NS. They used \swift{}/BAT, \maxi{}/GSC, \chandra{}/ACIS, and \swift{}/XRT data to study the spectral changes during the outburst, identifying a clear hard-to-soft state transition. Analysis of the archival \chandra{}/ACIS observations showed evidence for variations in the nonthermal component but not the thermal component during quiescence \citep{2014ApJ...780..127B}. \\

During the new 2023 outburst cycle of the source, \nicer{} collected $\sim 7.3$ ks of pointed observations starting on February 28, 2023, \citep{2023ATel15922....1S}. The source light curve shows a gradually increasing count rate, rising from $\sim 160$ counts s$^{-1}$ up to $\sim 230$ counts s$^{-1}$ in the $0.5 - 10$ keV energy range. \citet{2023ATel15922....1S} also performed spectral analysis of the persistent emission with $\sim 2.4$ ks exposure of \nicer{} data using an absorbed disk blackbody plus blackbody model in the 1-10 keV range. They reported the best-fit temperatures of the disk blackbody and blackbody components of $\sim 1 \kev{}$ and $\sim 2 \kev{}$, respectively. They derived an unabsorbed flux in the $0.5-10$ keV range of $2.61\times 10^{-9}$ erg~s$^{-1}$ cm$^{-2}$. Two Type-I X-ray bursts from this source were also detected with the \nicer{} during the early rising phase of the current outburst. Both Type-I bursts displayed a fast rise over a few seconds and exponential decay over $\sim 100$ seconds \citep{2023ATel15922....1S}. The burst decay is significantly longer than the previously observed burst from this source. A single Type-I X-ray burst (total duration $\sim 50$ seconds) from the source was also detected with a 10.6 ks \chandra{} observation on March 19, 2023  \citep{2023ATel15953....1H}. \nustar{} observed the source  on March 4 2023 which is $\sim 6$ days after its detection in this new outburst cycle. This \nustar{} observation allows us to perform a detailed spectral study of this transient source. \\

\subsection{IGR J17511-3057}
The source IGR J17511-3057 was discovered by \textit{INTEGRAL} through the Galactic Bulge monitoring program on September 12, 2009 \citep{2009ATel.2196....1B}. The position of the source was determined by \citet{2009ATel.2215....1N} using the \textit{Chandra}-High Energy Transmission Grating (\textsc{HETG}) observations to be at $\alpha_{J2000}=17^h51^m08^s.66$, $\delta_{J200}=-30^{\circ}57'41''.00$ (with $90 \%$ uncertainty of $0.6''$) and it was named IGR~J17511-3057. Just after its discovery, X-ray pulsations were detected at $245$ Hz by \rxte{} PCA bulge scans \citep{2009ATel.2197....1M}, which provided the NS spin frequency. Further, type-I X-ray bursts were reported by \citet{2009ATel.2198....1B}. Additionally, burst oscillations at $245$ Hz were also reported by \citet{2009ATel.2199....1W}. After its discovery, 18 type-I X-ray bursts were observed from this pulsar \citep{2010MNRAS.409.1136A, 2010A&A...509L...3B, 2010MNRAS.407.2575P, 2011A&A...529A..68F}. \citet{2010MNRAS.409.1136A} reported an upper limit to the distance of $6.96 \ \textrm{kpc}$ from the spectral properties of type-I X-ray bursts. The measured mass function indicates a main-sequence companion star with a mass around $0.13 M_\odot$ \citep{2009ATel.2197....1M}. \citet{2011ApJ...729....9K} reported on detecting the tentative twin kHz quasi-periodic oscillations (QPOs) exhibited by the source at the beginning and the end of the outburst. The separation between the QPOs is 120 Hz, around half of the NS spin frequency (245 Hz). \\

\citet{2010MNRAS.407.2575P} presented a detailed analysis of the X-ray spectra of IGR~J17511-3057 using 70 ks of \xmm{} data. They found that the spectrum of IGR~J17511-3057 is energetically dominated by a power law and detected signatures of reflection, such as a broadened iron K$\alpha$ emission line and a Compton hump at $\sim 30$ keV. They estimated an inner disc radius of $\sim 25\:R_{g}$ and an inclination within $38-68^{\circ}$ from the reflection component. \citet{2011A&A...529A..68F} also performed the spectral and timing analysis for the source using the simultaneous \inte{}, \rxte{} and \swift{} data obtained in September 2009 in the $0.8-300$ keV energy band. They reported that the broadband spectrum is well described by thermal Comptonization with an electron temperature of  $\sim 25$ keV, soft seed photon temperature of $kT_{bb}\sim 0.6\kev{}$ and Thomson optical depth $\tau_{T} \sim 2$ in a slab geometry. \citet{2012ApJ...755...52P} reported on the \chandra{}/{HETG} observation performed on September 22, 2009. The persistent emission is well described by the thermal Comptonization model of soft seed photons upscattered by a hot corona. They also detected a long ($\sim 54$ s) type-I X-ray burst and the spectrum of which can be well described with a blackbody with $kT_{bb} \sim 1.6 \kev{}$ and $R_{bb} \sim 5 \ \textrm{km}$. \\

\citet{2016A&A...596A..71P} reported a new outburst that occurred between March 23 and April 25, 2015, observed by \inte{}, \swift{} and \xmm{} \citep{2015ATel.7275....1B, 2015ATel.7288....1B}. During this outburst, the source reached a peak flux of $0.7 \times 10^{-9}$ erg cm$^{-2}$ s$^{-1}$ and decayed to quiescence in approximately a month. They found that the X-ray spectrum was dominated by a power law with a photon index between 1.6 and 1.8,  interpreted as the thermal Comptonized in an electron cloud with a temperature greater than 20 keV. Broad emission line ($\sigma=1 \ \textrm{keV}$) was also detected by them at E=$6.9^{+0.2}_{-0.3} \kev{}$, compatible with the K-$\alpha$ iron emission line. They further mentioned that the outburst flux and spectral properties shown during this outburst were remarkably similar to those observed during the previous accretion event detected from the source in 2009. Here, we present a detailed spectral analysis of the 2015 outburst of the accreting millisecond pulsar IGR~J17511-3057 as seen by \nustar{}.\\

\begin{table*}
\begin{center}
\centering 
\caption{List of \nustar{} observations.} 
\label{table:1}
\begin{tabular}{@{}lccccc}
\hline
Source  & Obs.\,ID & Obs.~Date & Instrument    & Net Exposure$^{*}$ & Count-rate \\
&  & & & (ksec) &(Count s$^{-1}$) \\
\hline 
Swift~J17480 &  80601304002 & March 4 2023  & FPMA \& FPMB  & $\sim 27.5$ &$\sim 170$  \\
IGR~J17511--3057 &  90101001002 & April 8 2015  & FPMA \& FPMB  & $\sim 42.2$ &$\sim 9$ \\
\hline
\end{tabular}
\end{center}
$^{*}$ The net exposures are based on per one telescope.
\end{table*}

This paper presents the spectral analysis of the source Swift~J17480 and IGR~J17511-3057 using the available \nustar{} observations. It is possible to constrain the accretion geometry of the system correctly by modeling the reflection spectrum using \nustar{} data because of its unprecedented sensitivity above 10 keV. Moreover, fitting with self-consistent reflection models will give essential insights, including the position of the magnetospheric radius and the neutron star magnetic field. We have organized the paper as follows: We give an overview of the observation and data reduction in Section 2. We present the source light curves in Section 3. We provide the details of the spectral analysis in Section 4. Finally, we discuss the results obtained from the analysis in Section 5. \\

\begin{figure*}
\centering
\includegraphics[width=0.45\textwidth, angle=0]{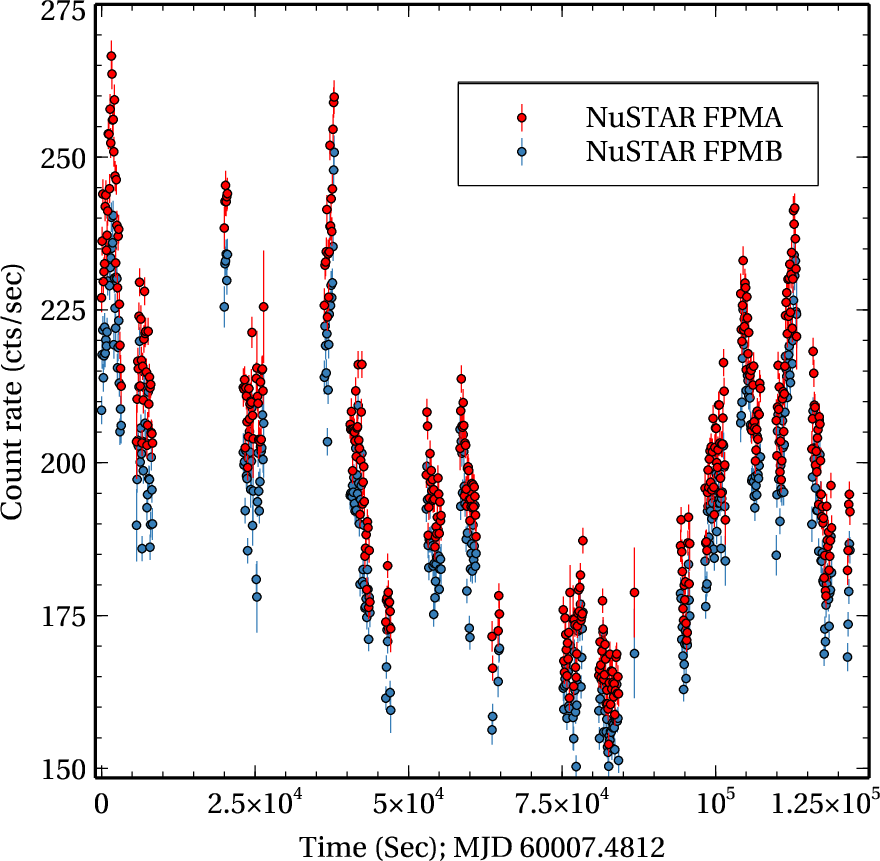}\hspace{1cm}
\includegraphics[width=0.45\textwidth, angle=0]{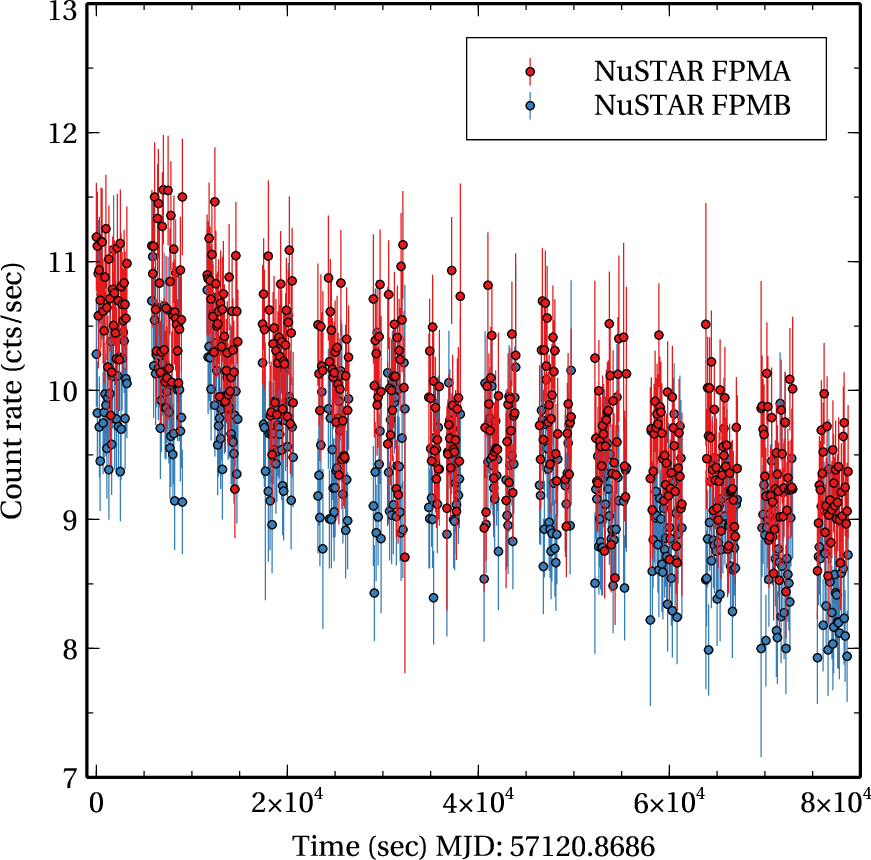}
\caption{Left panel: $3-40\kev{}$ \nustar{} FPMA (red) and FPMB (blue) light curves of the source Swift~J17480 with a binning of 100 s. Right panel: $3-79\kev{}$ \nustar{} FPMA (red) and FPMB (blue) light curves of the source IGR~J17511-3057 with a binning of 100 s.} 
\label{Fig1}
\end{figure*}

\section{Observation and Data reduction}
The \textit{Nuclear Spectroscopic Telescope ARray} (\nustar{}; \citealt{2013ApJ...770..103H}) observed the source Swift~J17480 only once on March 4, 2023, for a total exposure of $\sim 28$ ks. The \nustar{} also observed IGR~J17511-3057 once on April 8, 2015, for a total exposure of $\sim 42$ ks. All the \nustar{} observation details are listed in Table~\ref{table:1}. We have used those observations for our present analysis.  \\

The following data reduction techniques have been employed for both sources. The \nustar{} data were collected in the $3-79$\kev{} energy band using two identical focal plane modules, FPMA and FPMB, which are arranged in parallel. We processed the data using the standard \nustar{} data analysis software {\tt NUSTARDAS v2.1.2} task included in {\tt HEASOFT v6.33.2}. We used the latest \nustar{} calibration files {\tt CALDB} version available ($v20240206$) during the analysis. We used the task {\tt nupipeline v0.4.9} to generate the calibrated and screened event files. We have extracted the source events from a circular region centered on the source position, with a radius of $120''$ for both focal plane module detectors FPMA and FPMB. To collect background events, we selected a similar-sized region from the same chip but in a position away from the source for both instruments to avoid stray light from a nearby source. The filtered event files, the background subtracted light curves, and the spectra for both detectors are extracted using the tool {\tt nuproducts}. Corresponding response files are also created as output of {\tt nuproducts}, which ensures that all the instrumental effects, including loss of exposure due to dead time, are correctly accounted for. There was one type-I X-ray burst in the \nustar{} observation of the source IGR~J17511-3057. We created good time intervals (GTIs) to eliminate the burst from the spectra of the persistent emission. We grouped the FPMA and FPMB spectral data with a minimum of $100$ counts per bin to apply $\chi^{2}$ statistics during the fits. Finally, we fitted spectra from detectors FPMA and FPMB simultaneously, leaving a floating cross-normalization constant to account for differences in the effective area calibration of the detectors.

\begin{figure*}
\centering
\includegraphics[width=0.60\textwidth, angle=-90]{fig13a.eps}\\[10pt]
\includegraphics[width=0.60\textwidth, angle=-90]{fig13b.eps}
\caption{Left: Soft (3-10 keV) and hard (10-20 keV) light curves of the source Swift~J17480 with a binning of 100 s are shown in the top and middle panel, respectively. The bottom panel shows the variation of the hardness ratio which is defined here as the $10-20$ \kev{} count rate divided by the $3-10$ \kev{} count rate with time. Right: Soft (3-10 keV) and hard (10-70 keV) light curves of the source IGR~J17511-3057 with a binning of 100 s are shown in the top and middle panel, respectively. Here also, the bottom panel shows the variation of the hardness ratio which is defined here as the $10-70$ \kev{} count rate divided by the $3-10$ \kev{} count rate with time.} 
\label{Fig2}
\end{figure*}

\section{Light Curve}
In the left panel of Figure~\ref{Fig1}, we show the $3-40\kev{}$ \nustar{} FPMA and FPMB light curves of Swift~J17480, which spans the whole observation with 100 s time bins. It shows a little variability in the count rate. The source is detected with an average count rate of $\sim 170-200$ counts s$^{-1}$ with a minimum count between $\sim 75$ and $80$ ks during this \nustar{} exposure. Type-I X-ray bursts were not detected in the light curve. To construct the hardness ratio (HR) diagram, we have further generated light curves in the energy bands $3-10\kev{}$ and $10-20\kev{}$. We presented the resulting diagram in the left panel of Figure~\ref{Fig2}. We found that the $3-10\kev{}$ count rate is much higher compared to the $10-20\kev{}$ count rate, indicating a soft spectral state. However, the HR remains nearly constant between $0.09-0.11$, and no noticeable change in the HR is observed from this observation (see Figure~\ref{Fig2}). \\

For the source IGR~J17511-3057, we presented the $3-79\kev{}$ light curve with a binning of 100 s in the right panel of Figure~\ref{Fig1}. No patterns of variability have been observed during the observation. However, one type-I X-ray burst was detected around 83 ks during the \nustar{} observation, which is not shown in the light curve. We also eliminated the burst interval from the spectra of the persistent emission. The average count rate of this source during the observation was $\sim 10$ counts s$^{-1}$. We also constructed the HR diagram using the light curves in the energy bands $3-10\kev{}$ and $10-70\kev{}$, which is shown in the right panel of Figure~\ref{Fig2}. We note that the HR ratio remains constant ($\sim 0.4$) as expected from the light curves.

\begin{figure*}
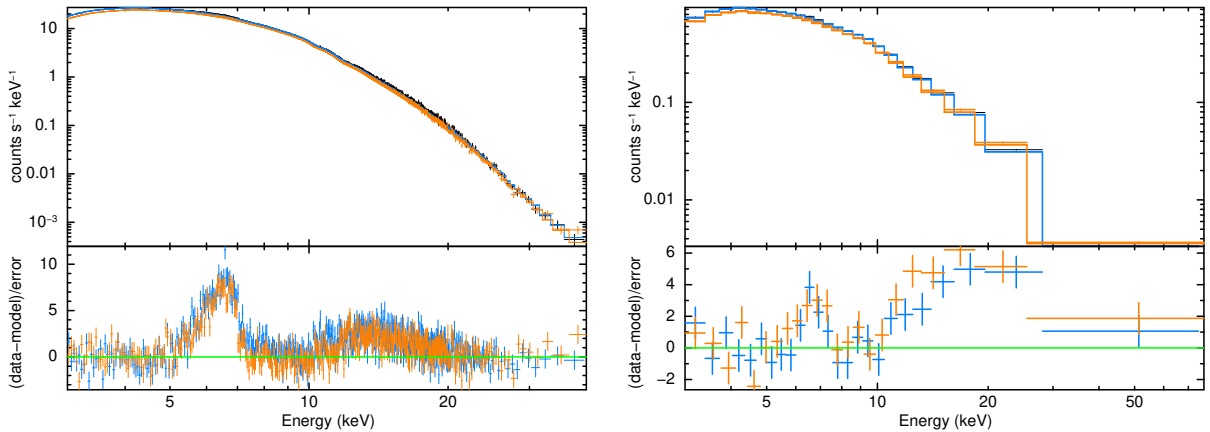

\includegraphics[scale=0.32, angle=-90]{fig14.eps}
\includegraphics[scale=0.32, angle=-90]{fig15.eps}
\caption{Left: The spectrum of the source Swift~J17480 in the energy band $3-40\kev{}$ obtained from the \nustar{} FPMA (orange) and FPMB (blue) is presented here. The continuum emission is fitted with two thermal components ({\tt diskbb} and {\tt bbody}) and a power-law component ({\tt powerlaw}). Right: The spectrum of the source IGR~J17511-3057 in the energy band $3-79\kev{}$ obtained from the \nustar{} FPMA and FPMB is shown here. The continuum emission is fitted with only absorbed cutoff power-law component ({\tt cutoffpl}). The spectra have been rebinned for plotting purposes. For both sources, the fit ratios associated with the continuum models are shown in the bottom panels of this plot. The presence of disc reflection is evident in the spectrum.} 
\label{Fig3}
\end{figure*}

\section{Spectral Analysis}
We used the X-ray spectral package {\tt XSPEC} $v12.14.0h$ \citep{1996ASPC..101...17A} to model the \nustar{} FPMA and FPMB spectra simultaneously between $3$ to $40$\kev{} energy band for the source Swift~J17480 and $3$ to $79$\kev{} energy band for the source IGR~J17511-3057. For the source Swift~J17480, we limited our analysis to the $3-40$\kev{} energy range as the background became significant at energies above $40\kev{}$. We included a multiplicative factor {\tt constant} that we fixed to one for the FPMA and left free for the FPMB to account for differences in the effective-area calibration of FPMA and FPMB. We used the {\tt TBabs} model to account for interstellar absorption along the line of sight with the {\tt wilm} abundances \citep{2000ApJ...542..914W} and the {\tt vern} \citep{1996ApJ...465..487V} photoelectric cross-section. The only free parameter of the {\tt TBabs} component is $N_{H}$, the column density of the absorbing component along the line of sight. The spectral uncertainties are reported at the $1 \sigma$ confidence level, unless otherwise stated.

\subsection{Swift~J17480}
Different combinations of model components and spectral parameters are required to describe the continuum emission from different spectral states from the NS LMXBs. To probe the spectral shape of this source correctly, we tried to fit the continuum emission using two thermal components consisting of an absorbed multi-temperature blackbody ({\tt diskbb},\citealt{1984PASJ...36..741M}) to model the emission from the accretion disc, and a single-temperature blackbody ({\tt bbody}) to model the emission from the NS surface or a region close to the NS. Because of the lack of sensitivity of \nustar{} at low energies, we had to fix the photoelectric equivalent hydrogen column density at $0.54 \times 10^{22}$ cm$^{-2}$, following \citet{1990ARA&A..28..215D}. We found that the fit with this two-component thermal model is statistically poor and leaves significant residuals associated with strong reflection features in the $5-9$ \kev{} and $12-30$ \kev{} energy range and with an excess at higher energies (above $30\kev{}$). The presence of this hard tail implies the need for an additional continuum component. The observed residual at high energy is related to the hard power-law tails which have been observed in many NS X-ray binary systems \citep{2000ApJ...544L.119D} as well as AMSPs (see e.g. \citealt{2016MNRAS.457.2988P}, and references therein). We added a {\tt powerlaw} (in {\tt xspec}) component to model the same. This component was able to flatten the higher energy residuals and was statistically significant ($\Delta\chi^2=-29$ for 2 additional dof). Although the model {\tt TBabs*(diskbb+bbody+powerlaw)} describes the continuum emission reasonably well, the poor quality of the fit ($\chi^2/dof=3251/955$) is mainly due to the presence of strong reflection features. We found an inner disc temperature of $kT_{in} \sim 1.8 \kev{}$, an NS surface or spreading layer (between the disk and the NS surface) temperature of $kT_{bb} \sim 2.5\kev{}$, and an asymptotic power-law photon index $\Gamma \sim 2.8$. These spectral parameters suggest the source was observed during a soft spectral state. We note that a continuum consisting of a disk blackbody, a blackbody, and a power law is also frequently used for soft state spectra of many NS LMXBs (e.g.,\citealt{2007ApJ...667.1073L, 2010ApJ...720..205C, 2013ApJ...779L...2M, 2019ApJ...873...99L})\\

It has been observed that for NS LMXBs in their soft states, the hard part of the X-ray spectrum is typically modeled as either a hot ($\simeq 2-3$ keV) blackbody or thermal Comptonization. Considering this, we employed a thermal Comptonization model {\tt nthcomp} \citep{1996MNRAS.283..193Z} instead of a blackbody. We set the photon seed input to a disc blackbody. However, residuals are still observed above $30\kev{}$, even after including the thermal Comptonization model. It suggests that the thermal Comptonization component alone cannot describe the hard X-ray emission, and a non-thermal component is necessary to model the high-energy tail. We found that the model {\tt TBabs*(diskbb+nthcomp+powerlaw)} fits our data less well with $\chi^2/dof=3256/953$. For the {\tt nthcomp} component, we obtained the electron temperature $kT_{e}\sim 3.0\kev{}$ with photon index $\Gamma\sim 1.88$. The seed photon temperature is found to be high $kT_{seed}\sim 1.9\kev{}$, comparable to the inner disk temperature $kT_{in}$. The spectral parameters also suggest a soft spectral state. The continuum emission may be physically interpreted as direct emission from the accretion disc and Comptonized emission from a boundary layer (or a hot flow) with the seed photons provided by the disc. However, our continuum fits prefer a black body over a Comptonized model, presumably because the Comptonized emission is well approximated by the hot black body as observed in soft spectral states of NS LMXBs \citep{2016MNRAS.461.4049D}. We note that the {\tt nthcomp} with seed photons provided by a blackbody returned some inconsistent, unacceptable results.\\

Both choices of our continuum modeling leaves large positive residuals around $5-8$\kev{} and $10-20$\kev{}, as shown in the left panel of Figure~\ref{Fig3}. It indicates the presence of a broad iron emission feature ($\sigma\sim 0.61$\kev{}, EW$\sim 105$ eV, measured using a {\tt gaussian} emission line) and Compton reflection hump, respectively. These features are associated with hard photons reflected from the disc. It suggests fitting the broadband spectrum with a self-consistent reflection model. Moreover, we note that the Compton hump is visible but not much prominent in the \nustar{} spectra. The above usually happens when the reflection is produced by a softer illuminating spectrum with respect to the typical non-thermal (power-law) continuum considered in standard reflection models. To take into account the reflected photons, we added {\tt relxillNS} \citep{2022ApJ...926...13G}, a variant of the model {\tt relxill} used to calculate the relativistic reflection from the innermost regions of the accretion disk (\citealt{2014ApJ...782...76G, 2014MNRAS.444L.100D}). In particular, {\tt relxillNS} assumes a single-temperature blackbody spectrum as the primary continuum that illuminates the disc, physically related to the NS surface or the spreading layer emission. In our case, the single-temperature blackbody component provides most of the hard X-ray flux that illuminates the accretion disc, consistent with the nature of {\tt relxillNS}. It motivated us to choose {\tt relxillNS} out of all the reflection models appropriate for an incident blackbody spectrum. Moreover, in NS systems, it has been observed in many cases that the emission from the NS surface/boundary layer is significant and contributes to the reflection \citep{2008ApJ...674..415C, 2019ApJ...873...99L, 2017ApJ...836..140L}. \\

The model {\tt relxillNS} has similar parameters to that of {\tt relxill} with the addition of log $n$ (cm$^{-3})$ to vary the density of the accretion disc and $kT_{bb}$(\kev{}) for the input blackbody spectrum instead of $\Gamma$ and $E_{cut}$. We tied the {\tt relxillNS} $kT_{bb}$ to the blackbody temperature of the {\tt bbody}. As we consider the illumination spectrum separately, we set reflection fraction $r_{refl}=-1$. We let the inner disc radius $R_{in}$, disc inclination $i$, ionization parameter log$\xi$, Fe abundance $A_{Fe}$, and electron density log$n_{e}$ free to vary during the fitting. We fixed the emissivity indices $q_{1}=q_{2}=3$ and the outer disc radius $R_{out}=1000\:R_{g}$ (where $R_{g}=GM/c^{2}$ is the gravitational radius). For a standard NS, the spin period $P$ can be used to derive the dimensionless spin parameter, adopting $a= 0.47/P(ms)$ \citep{2000ApJ...531..447B}. However, the spin period of the source Swift~J17480 is not known. We therefore performed the fit with $a=0$, corresponding to non-spinning NS and $a=0.3$, corresponding to fastest known spinning NS at $\sim 1.5$ ms \citep{2008ApJS..179..360G}. The addition of this reflection component provided excellent fits to the \nustar{} spectra with $\chi^2/dof=1100/949$ ($\Delta\chi^2=-2151$ for 6 additional dof). The fit was able to constrain all the free parameters. We note that fits performed with $a=0$ yielded similar results as for $a=0.3$. The best-fitting parameters are reported in Table~\ref{table:2} while the spectra are represented in Figure~\ref{Fig4}. We finally tested the goodness of the fit for the inner disk radius, $R_{in}$, and disc inclination angle, $i$. We used the command {\tt steppar} in {\tt xspec} to search the best-fit for $R_{in}$ and $i$ for the best-fit model {\tt const*TBabs*(diskbb+bbody+powerlaw+relxillNS)}. The left and right panels of Figure~\ref{Fig6} show the variation of $\Delta\chi^{2}(=\chi^{2}-\chi_{min}^{2})$ as a function of $R_{in}$ and $i$, respectively for the best-fit model. \\

\begin{figure*}
\centering
\includegraphics[scale=0.34, angle=-90]{fig16.eps}
\caption{Swift~J17480: \nustar{} unfolded spectra and model components fitted with the best-fit model {\tt const*TBabs*(diskbb+bbody+powerlaw+relxillNS)}. The lower panel of this plot shows the ratio of the data to the model in units of $\sigma$.}
\label{Fig4}
\end{figure*}

\subsection{IGR~J17511-3057}
We tried to fit the continuum emission (\nustar{} spectrum $3-79$\kev{}) with a model consisting of an absorbed cutoff power-law ({\tt cutoffpl}) only. The column density along the line of sight is fixed at $N_{H}=0.53\times 10^{22}$ cm$^{-2}$. This model alone describes the continuum emission reasonably well with $\chi^2/dof=1135/1079$. The spectrum is dominated by a hard Comptonization component from the corona with a photon index of $\Gamma=1.70\pm 0.01$ and the cutoff energy $E_{cut}= 95\pm 4$\kev{}, consistent with \citet{2011A&A...529A..68F} and \citet{2016A&A...596A..71P}. There is a broad asymmetric Fe line feature and Compton hump present in the ratio of the data to the continuum (see right panel of Figure~\ref{Fig3}). A {\tt gaussian} emission line component measured $\sigma\sim 0.32$\kev{} and EW$\sim 25$ eV for the Fe emission line. Although the detection of the Fe line component is not a strong one, the Compton reflection hump is clearly visible. This may be because of the strong Comptonization and geometry of the system \citep{2019ApJ...873...99L}. \\

\begin{figure*}
\centering
\includegraphics[scale=0.34, angle=-90]{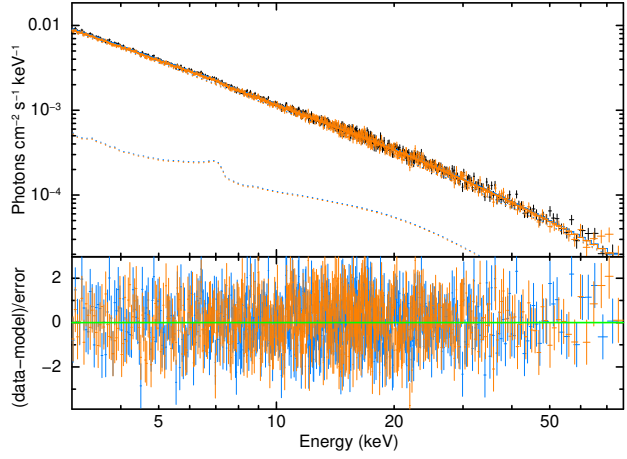}
\caption{IGR~J17511-3057: \nustar{} unfolded spectra and model components fitted with the best-fit model {\tt const*TBabs*(cutoffpl+relxill)}. The lower panel of this plot shows the ratio of the data to the model in units of $\sigma$.} 
\label{Fig5}
\end{figure*}

\begin{figure*}
\centering
\includegraphics[scale=0.40, angle=0]{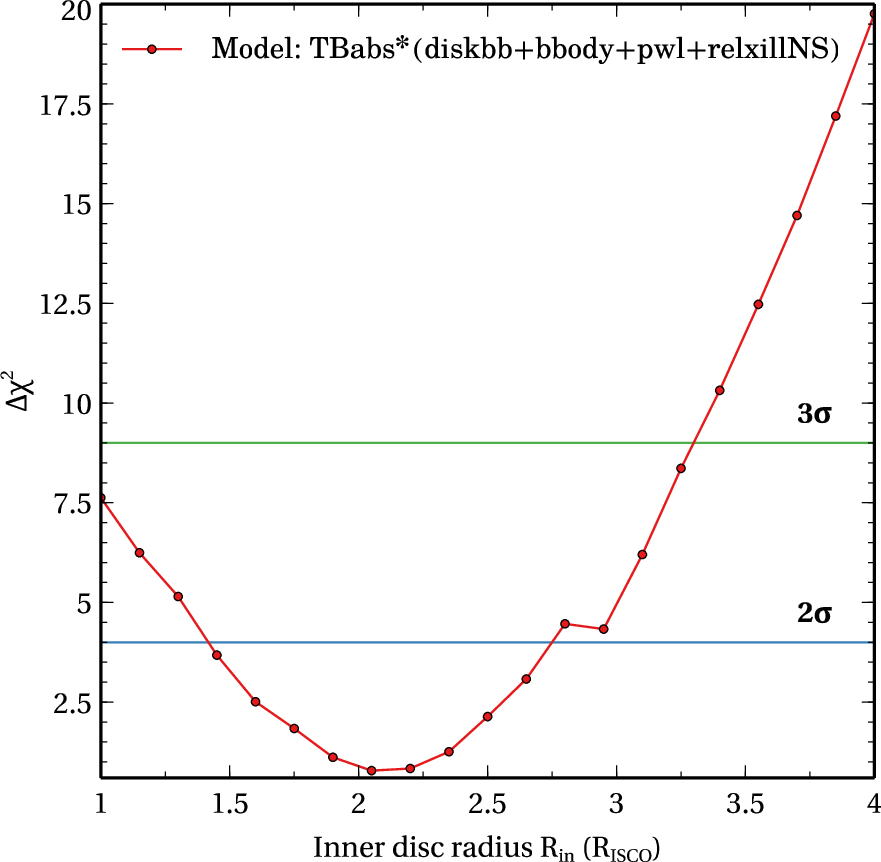}\hspace{2cm}
\includegraphics[scale=0.40, angle=0]{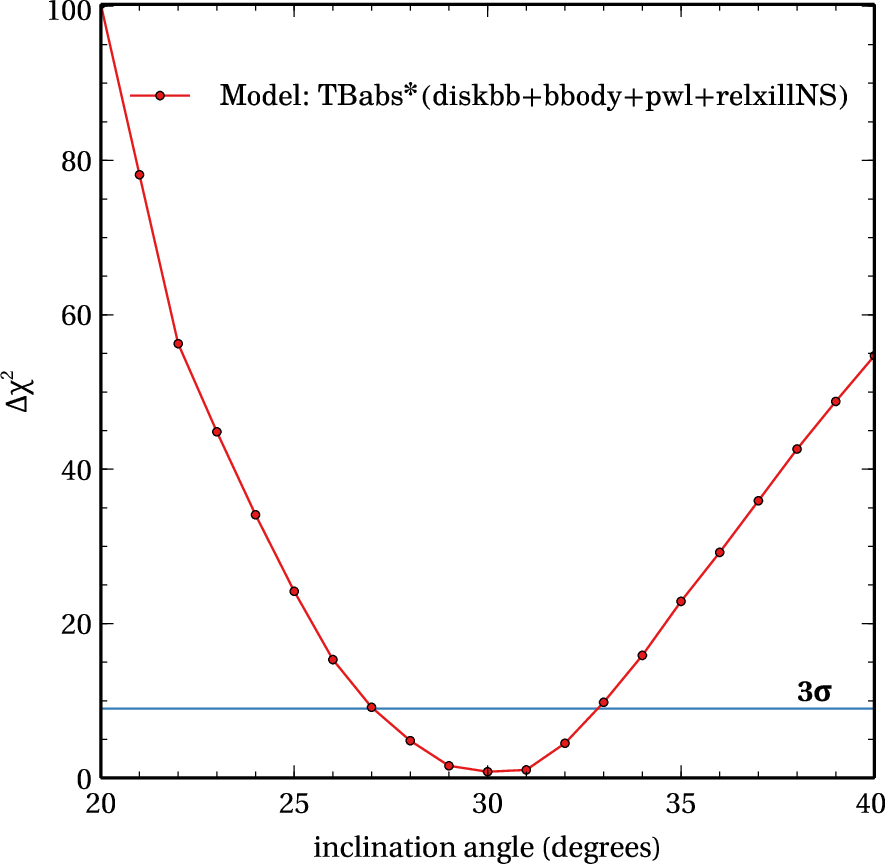}
\caption{Source Swift~J17480: The plots show the change in the goodness of fit for the inner disc radius ($R_{in}$) and disc inclination angle ($i$). The left and right panel show the variation of $\Delta\chi^{2}(=\chi^{2}-\chi_{min}^{2})$ as a function of $R_{in}$ (varied between $1$ to $4\:R_{ISCO}$) and $i$ (varied between $20\degree$ to $40\degree$ ), respectively, obtained from the best-fit model.} 
\label{Fig6}
\end{figure*}

\begin{figure*}
\centering
\includegraphics[scale=0.40, angle=0]{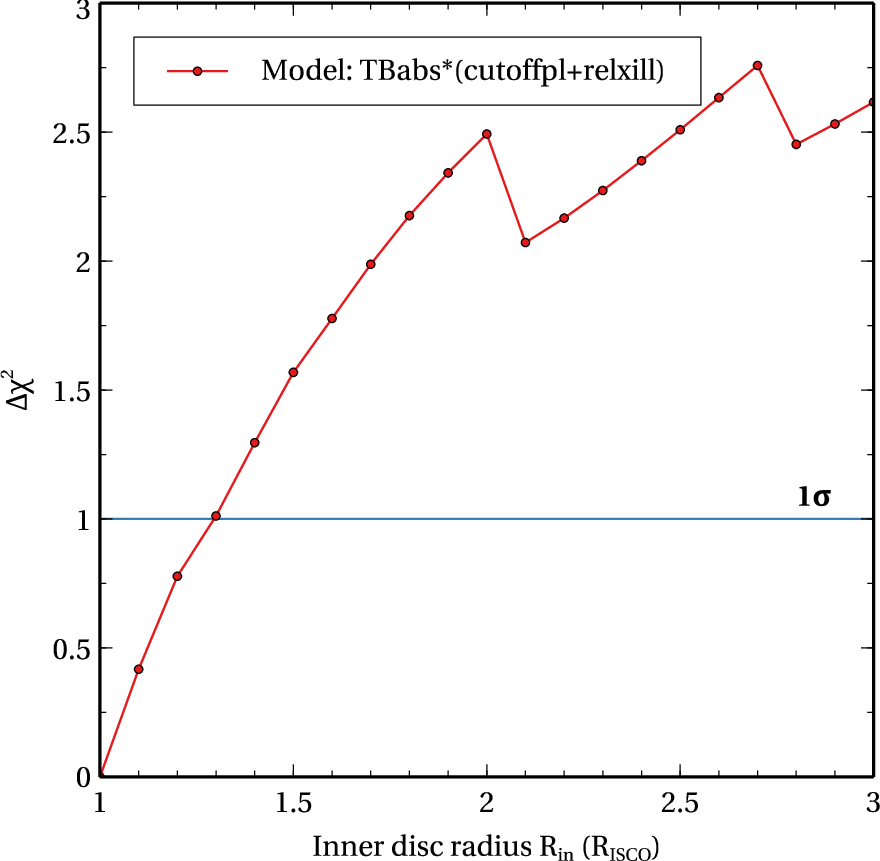}\hspace{2cm}
\includegraphics[scale=0.40, angle=0]{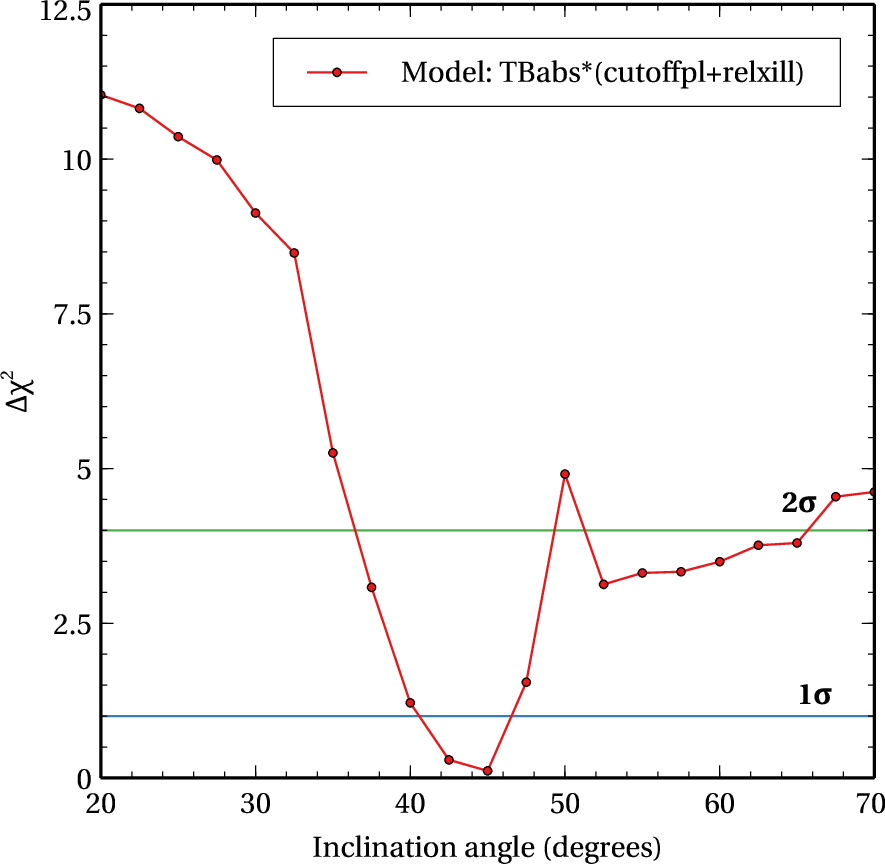}
\caption{Source IGR~J17511-3057: The plots show the change in the goodness of fit for the inner disc radius ($R_{in}$) and disc inclination angle ($i$). The left and right panel show the variation of $\Delta\chi^{2}(=\chi^{2}-\chi_{min}^{2})$ as a function of $R_{in}$ (varied between $1$ to $3\:R_{ISCO}$) and $i$ (varied between $20\degree$ to $70\degree$ ), respectively, obtained from the best-fit model.} 
\label{Fig7}
\end{figure*}

In order to accommodate these features, we applied the standard version of the self-consistent reflection model {\tt relxill}. We fixed the emissivity indices $q_{1}=q_{2}=3$, the outer disc radius $R_{out}=1000\:R_{g}$, and $r_{refl}=-1$. We further fixed the dimensionless spin parameter at $a=0.115$ as the source exhibits coherent pulsation at $4.1$ ms. We tied the {\tt relxill} $\Gamma$ and $E_{cut}$ to the {\tt cutoffpl} $\Gamma$ and $E_{cut}$. Allowing the spectrum to be described by reflection with {\tt relxill} provides an improvement to the fit leading to $\chi^2/dof=1057/1074$ ($\Delta\chi^2=-78$ for 5 additional dof). Most of the free parameters are well constrained and reported in Table~\ref{table:3}. The spectrum and the spectral components can be seen in Figure~\ref{Fig5}. However, if we continue the spectral fitting keeping $r_{refl}$ free, the self-consistent reflection model allows us to find that the $r_{refl}$ is $0.25_{-0.05}^{+0.03}$ which means less number of photons interacted with the surrounding disc first rather than emitted directly to the observer, as evident from the reflection features. Here also, the goodness of the fit has been tested for the inner disk radius, $R_{in}$, and disc inclination angle, $i$ using the {\tt steppar} command in {\tt xspec} for the best-fit model {\tt const*TBabs*(cutoffpl+relxill)}. The variation of $\Delta\chi^{2}(=\chi^{2}-\chi_{min}^{2})$ as a function of $R_{in}$ and $i$ for the best-fit model are shown in the left and right panel of Figure~\ref{Fig7}. \\

 \begin{table*}
   \centering
\caption{Fit results: Best-fitting spectral parameters of the \nustar{} observation of the source Swift~J17480 using the model
    {\tt const*TBabs*(diskbb+bbody+powerlaw+relxillNS)}}.
\begin{tabular}{|p{3.4cm}|p{5.5cm}|p{3.2cm}|}
    \hline
    Components     & Parameters (unit) & Best-fit values \\
    \hline
    {\scshape Constant} & FPMB (wrt FPMA) & $0.99\pm 0.003$ \\
    {\scshape tbabs}    & $N_{H}$($\times 10^{22}\;\text{cm}^{-2}$) & $0.54 $(f)   \\
    {\scshape diskbb} & $kT_{in} (\kev)$ & $1.82 \pm 0.05$    \\
    & Norm & $35_{-1}^{+3}$  \\
    {\scshape bbody} & $kT_{bb} (\kev)$ & $2.47_{-0.07}^{+0.04}$   \\
    & Norm ($\times 10^{-2}$)& $1.38_{-0.14}^{+0.09}$  \\
    {\scshape powerlaw} & $\Gamma$ & $2.57_{-0.20}^{+0.15}$  \\
    & Norm & $0.12_{-0.06}^{+0.08}$    \\  
    {\scshape relxillNS} & $i$ (degrees) & $30 \pm 1$  \\
    & $R_{in}$ ($\times R_{ISCO}$) & $2.36_{-0.37}^{+0.32}$  \\
    & $kT_{bb} (\kev)$ & = blackbody $kT_{bb}$   \\
    & $\rm{log}\:\xi$ (erg cm s$^{-1}$) &  $2.70_{-0.04}^{+0.02}$ \\
    & $A_{Fe}$ ($\times \;\text{solar})$   & $1.04_{-0.10}^{+0.35}$ \\
    & log$n_{e}$ (cm$^{-3}$) &  $17.01_{-0.63}^{+0.93}$  \\
    & $f_{refl}$   & $-1$(f)  \\
    & norm ($\times 10^{-3}$)   &  $1.59_{-0.09}^{+0.11}$ \\
    {\scshape cflux} & $F_{diskbb}^{*}$ ($\times 10^{-9}$ ergs/s/cm$^2$) & $4.0\pm 0.03$\\
    & $F_{bbody}^{*}$ ($\times 10^{-9}$ ergs/s/cm$^2$) & $1.5\pm 0.02$\\
    & $F_{powerlaw}^{*}$ ($\times 10^{-9}$ ergs/s/cm$^2$) & $0.14\pm 0.01$\\
    & $F_{relxillNS}^{*}$ ($\times 10^{-9}$ ergs/s/cm$^2$) & $0.66\pm 0.01$ \\
    & $F_{total}^{*}$ ($\times 10^{-9}$ ergs/s/cm$^2$) & $6.3\pm 0.01$ \\
   \hline 
    & $\chi^{2}/dof$ & $1100/949$  \\
    \hline
  \end{tabular}\label{table:2} \\
{\bf Note:} The outer radius of the {\tt relxillNS} spectral component was fixed to $1000\;R_{g}$. We fixed emissivity index $q1=q2=3$ (\citealt{2010ApJ...720..205C, 2019ApJ...873...99L}). The spin parameter ($a$) was fixed to $0.005$ and the redshift was set to to zero for the {\tt relxillNS} model. $^{*}$All the unabsorbed fluxes are calculated in the energy band $3-40 \kev{}$ using the {\tt cflux} model component in {\tt XSPEC}. \\

\end{table*}

\begin{table*}
   \centering
\caption{Fit results: Best-fitting spectral parameters of the \nustar{} observation of the source IGR~J17511-3057 using the model
    {\tt const*TBabs*(cutoffpl+relxill)}}.
\begin{tabular}{|p{3.4cm}|p{5.5cm}|p{2.8cm}|}
    \hline
    Components     & Parameters (unit) & Best-fit values \\
    \hline
    {\scshape Constant} & FPMB (wrt FPMA) & $0.97\pm 0.005$ \\
    {\scshape tbabs}    & $N_{H}$($\times 10^{22}\;\text{cm}^{-2}$) & $0.53 $(f)   \\
    {\scshape cutoffpl} & $\Gamma$ & $1.73 \pm 0.01$  \\
    & $E_{cut} (\kev)$ & $95_{-5}^{+4}$ \\
    & Norm ($\times 10^{-2}$) & $6.5_{-0.12}^{+0.07}$    \\  
    {\scshape relxill} & $i$ (degrees) & $44_{-4}^{+3}$  \\
    & $R_{in}$ ($\times R_{ISCO}$) & $\lesssim 1.3$  \\
    & $\Gamma$ & = cutoffpl $\Gamma$ \\
    & $\rm{log}\:\xi$ (erg cm s$^{-1}$) &  $2.68_{-0.08}^{+0.06}$ \\
    & $A_{Fe}$ ($\times \;\text{solar})$   & $\lesssim 0.54$ \\
    & $E_{cut} (\kev)$ & = cutoffpl $E_{cut}$ \\
    & $f_{refl}$   & $-1$(f)  \\
    & norm ($\times 10^{-4}$)   &  $2.92_{-0.46}^{+0.35}$ \\
    {\scshape cflux} & $F_{cutoffpl}^{*}$ ($\times 10^{-10}$ ergs/s/cm$^2$) & $5.50\pm 0.03$\\
    & $F_{relxill}^{*}$ ($\times 10^{-10}$ ergs/s/cm$^2$) & $0.78\pm 0.02$ \\
    & $F_{total}^{*}$ ($\times 10^{-10}$ ergs/s/cm$^2$) & $6.28\pm 0.01$ \\
   \hline 
    & $\chi^{2}/dof$ & $1057/1074$  \\
    \hline
  \end{tabular}\label{table:3} \\
{\bf Note:} The outer radius of the {\tt relxill} spectral component was fixed to $1000\;R_{g}$. We fixed emissivity index $q1=q2=3$ (\citealt{2010ApJ...720..205C, 2019ApJ...873...99L}). The spin parameter ($a$) was fixed to $0.115$ as the source is detected with a coherent pulsation at $4.1$ ms. $^{*}$All the unabsorbed fluxes are calculated in the energy band $3-79 \kev{}$ using the {\tt cflux} model component in {\tt XSPEC}. \\

\end{table*}

\begin{table}
\centering
\caption{Estimation of various parameters.} 
\begin{tabular}{|p{2.5cm}|p{3.0cm}|p{3.0cm}|}
\hline
Parameters (Unit) & Swift~J17480 & IGR~J17511-3057 \\  
\hline
$R_{in}$ ($ R_{g}$) &   $11.94-16.1$ &  $\lesssim 7.32$  \\

$\dot{m}$ ($10^{-9}\;M_{\odot}\;\text{y}^{-1}$) & $\sim 1.7$ & $\sim 0.56$ \\

$R_{BL, max} (R_{g})$ & $\sim 12.1$ & $\sim 5.9$ \\

$B_{max}$ ($ 10^{9}$ G) &  $\le 1.0$ &  $\le 0.096$\\

$R_{co}$ (km)  & $ -- $  &  $\sim 42$ \\
\hline
\end{tabular}\label{table:4} \\
Where $R_{g}=GM/c^{2}$ is the gravitational radius and $R_{g}=2.1$ km for NS of 1.4 $M_{\odot}$. Precise estimation of $R_{co}$ is not possible for the source Swift~J17480 as the spin period of the source is unknown. See text (mainly discussion section) for the details of the calculations of the parameters.
\end{table}

\section{Discussion}
We present the results of the spectral analysis of the NS LMXB Swift~J17480 and AMSP IGR~J17511-3057 using the \nustar{} observations performed on March 4, 2023, and April 8, 2015, respectively. The source Swift~J17480 is detected with a high average count rate of $\sim 170-200\ctssec$ during this \nustar{} observation. The unabsorbed $3-40\kev{}$ flux is estimated at $6.3\times 10^{-9}$ erg~s$^{-1}$ cm$^{-2}$ which is $\sim 2$ times larger than the \nicer{} flux in the $0.5-10$\kev{} energy band observed on February 28, 2023, one day after its discovery \citep{2023ATel15922....1S}. The unabsorbed bolometric X-ray flux during this observation in the energy band $0.1-100\kev{}$ is $8.84\times 10^{-9}$ erg~s$^{-1}$ cm$^{-2}$, implying an unabsorbed bolometric luminosity of $3.68\times 10^{37}$ erg~s$^{-1}$, assuming a distance of $5.9$ kpc. This value corresponds to $\sim 10\%$ of the Eddington luminosity ($L_{Edd}$) which is $\sim 3.8\times 10^{38}$ erg~s$^{-1}$ for a canonical $1.4\:M_{\odot}$ NS \citep{2003A&A...399..663K}. Whereas, the source IGR~J17511-3057 is detected with an average count rate of $\sim 10-11\ctssec$, resulting an unabsorbed $3-79\kev{}$ flux of $6.28\times 10^{-10}$ erg~s$^{-1}$ cm$^{-2}$, consistent with the peak flux during 2015 outburst \citep{2016A&A...596A..71P}. The unabsorbed bolometric luminosity of the source is $4.46\times 10^{36}$ erg~s$^{-1}$, corresponds to only $\sim 1\%$ of $L_{Edd}$, assuming a distance of $6.9$ kpc. \\

We found that the persistent X-ray emission of the source Swift~J17480 is well described with a combination of two soft thermal components ({\tt diskbb} and {\tt bbody}) and a power-law to account for the observed high energy tail above $30\kev{}$. The continuum components are associated with emission from the accretion disc, the NS surface and/or boundary layer (between the disk and the NS surface). Moreover, the continuum emission can also be described using a soft thermal component, a thermal Comptonization component, and a power-law emission. It may be physically interpreted as a direct emission from the accretion disc and a Comptonized emission from a boundary layer (or a hot flow) with the seed photons provided by the disc. We found that the Comptonized emission is well approximated by a hot black body, as is typically the case in soft spectral states. However, the detection of hard X-ray power-law may be associated with either emission of Comptonization by a non-thermal medium \citep{1998PhST...77...57P} or bulk motion of accreting material close to the NS \citep{2000ApJ...544L.119D}. A non-thermal emission from this source was also predicted by \citet{2014ApJ...780..127B} during the 2012 outburst of the source. The inferred spectral parameters from both continuum models suggest that the source was observed during a soft spectral state. The spectral nature is very similar to the last outburst that occurred in 2012, as a clear hard-to-soft state transition was identified during the outburst \citep{2014ApJ...780..127B}. The spectrum exhibited the presence of a broad ($\sigma\sim 0.61$\kev{}) Fe-K emission line in $5-8\kev{}$, a Compton hump $10-20\kev{}$, indicating the reflection from the inner accretion disc. Whereas, the continuum emission of the source IGR~J17511-3057 is well described by a hard Comptonization component with power-law $\Gamma \sim 1.71$ and a cut-off energy $ \sim 95\kev{}$, suggesting a behavior similar to the most of AMSPs (\citealt{2020A&A...641A..37K, 2021A&A...649A..76L, 2022MNRAS.516L..76S, 2023MNRAS.519.3811S, 2023ApJ...958..177L}). We found clear evidence of disc reflection of the coronal emission from the \nustar{} spectrum of IGR~J17511-3057.\\

The observed features of the continuum modeling motivate us to employ the disc reflection models for both sources. For the source Swift~J17480, we employed a reflection component {\tt relxillNS} for which it is assumed that a single-temperature blackbody spectrum is irradiating the disc, physically related to the NS surface or the spreading layer emission. We found that the $3-40\kev{}$ source energy spectrum is adequately fitted using a model combination consisting of two thermal components ({\tt diskbb} and {\tt bbody}), a power-law component ({\tt powerlaw}) and a reflection model {\tt relxillNS}. From the disc reflection modeling, we found that the inner disc radius is very precisely constrained to $R_{in}=(1.99-2.68)\:R_{ISCO}=(11.94-16.1)R_{g}$ ($25.1-33.8$ km for a $1.4\Msun$ NS), where $R_{ISCO}=6\:R_{g}$ for a non-spinning NS with $a=0$, which is consistent with the other Terzan 5 source IGR~J17480-2446 \citet{2011ApJ...731L...7M}. The reflection fit yielded an inclination estimate of $i=30\degree\pm 1$, consistent with no dips or eclipses observed in this source's light curve. Moreover, a low system inclination ($\sim 20\degree$) was also estimated by \citet{2011ApJ...731L...7M} for the source IGR~J17480-2446 located at Terzan 5. The reflection fit also revealed a moderate disc ionization parameter with log$\xi=2.70_{-0.04}^{+0.02}$, consistent with the typical range observed in both black hole and NS LMXBs (log($\xi$) $\sim 2-3$). The iron abundance is also consistent with the solar composition $A_{Fe}=1.04_{-0.10}^{+0.35}$. The density of the disc is estimated to be log$(n_{e}/\rm{cm}^{-3})=17.01_{-0.63}^{+0.93}$. The inferred disc radius $R_{in}$ and inclination $i$ of the source Swift~J17480 is consistent with other NS LMXBs (\citealt{2010ApJ...720..205C, 2015MNRAS.451L..85D, 2015MNRAS.449.2794D, 2016ApJ...824...37L, 2017MNRAS.466.4991M}). \\

For the source IGR~J17511-3057, we employed the standard relativistic reflection model {\tt relxill}. The $3-79\kev{}$ source energy spectrum is well fitted with a combination of {\tt cutoffpl} and {\tt relxill}. We found an $1 \sigma$ upper limit to the inner disc radius of $\sim 1.3\:R_{ISCO}=7.32\:R_{g}$ ($15.5$ km for a $1.4\Msun$ NS) where $R_{ISCO}=5.63\:R_{g}$ for a spinning NS with $a=0.115$. 
The inferred value of $R_{in}$ is somewhat smaller, which is also predicted by \citet{2016A&A...596A..71P} based on \inte{}, \swift{}, and \xmm{} observations of IGR~J17511-3057 performed during the same outburst cycle that occurred between March 23 and April 25, 2015. They obtained a value of $R_{in}<\:8.5\:R_{g}$ using a {\tt diskline} component. Motivating with this, they fixed the value of $R_{in}$ to $10\:R_{g}$ during the fitting with the {\tt reflionx} model. Although the estimation of $R_{in}$ using \nustar{} observation is in line with that of \swift{} and \xmm{} observation, it may be accepted with a degree of caution as the \nustar{} spectrum allowed us to weakly constrained the value of $R_{in}$. The inferred inclination angle of the system is $i=44_{-4}^{+3}$$\degree$, perfectly consistent with \citet{2010MNRAS.407.2575P}. The {\tt relxill} model yields a moderate disc ionization, log$\xi=2.68_{-0.08}^{+0.06}$ and low iron abundance, $A_{Fe}\lesssim 0.54$, consistent with the observed reflection continuum. Moreover, for the pulsars with period $P$ (ms), the co-rotation radius can be estimated using $R_{co}=(GMP^{2}/4\pi^{2})^{1/3}$. For the source IGR~J17511-3057, spinning at $4.1$ ms, the value of $R_{co}$ is estimated to be $\sim 42$ km ($20\:R_{g})$. Our estimate of $R_{in}$ for the source IGR~J17511-3057 is consistent, considering that the accretion disc cannot be truncated outside the co-rotation radius ($R_{in}\lesssim R_{co}$). However, for the source Swift~J17480, the estimation of $R_{co}$ is not possible without precisely knowing its spin period. \\

Our best-fit spectral parameters have been used to compute physical properties like mass accretion rate ($\dot{m}$), the maximum radius of the boundary layer ($R_{BL}$), and magnetic field strength ($B$) of the NS in the system. We first estimated the mass accretion rate per unit area, using Equation (2) of \citet{2008ApJS..179..360G}
\begin{equation}
\begin{split}
\dot{m}=&\:6.7\times 10^{3}\left(\frac{F_{p}\:c_\text{bol}}{10^{-9} \text{erg}\: \text{cm}^{-2}\: \text{s}^{-1}}\right) \left(\frac{d}{10 \:\text{kpc}}\right)^{2} \left(\frac{M_\text{NS}}{1.4 M_{\odot}}\right)^{-1}\\
 &\times\left(\frac{1+z}{1.31}\right) \left(\frac{R_\text{NS}}{10\:\text{km}}\right)^{-1} \text{g}\: \text{cm}^{-2}\: \text{s}^{-1}.
 \end{split} 
\end{equation}
The above equation yields $\dot{m}$ of $1.7\times 10^{-9}\;M_{\odot}\;\text{y}^{-1}$ at a persistent flux $F_{p}=6.3\times 10^{-9}$ erg~s$^{-1}$ cm$^{-2}$ for the source Swift~J17480 and $\dot{m}$ of $5.6\times 10^{-10}\;M_{\odot}\;\text{y}^{-1}$ at a persistent flux $F_{p}=6.28\times 10^{-10}$ erg~s$^{-1}$ cm$^{-2}$ for the source IGR~J17511-3057. Here, we assume the bolometric correction $c_{bol} \sim 1.38$ \citep{2008ApJS..179..360G} and $1+z=1.31$ (where $z$ is the surface redshift) for an NS with mass ($M_{NS}$) 1.4 $M_{\odot}$ and radius ($R_{NS}$) $10$ km. We further used Equation (2) of \citet{2001ApJ...547..355P} to estimate the maximum radial extension of the boundary layer ($R_{BL}$) from the NS surface based on the mass accretion rate. We found the maximum value of $R_{BL}$ to $\sim 12.1\;R_{g}\: (2.02 \:R_{ISCO})$ for the source Swift~J17480 and $\sim 5.9\;R_{g}\: (1.05 \:R_{ISCO})$ for the source IGR~J17511-3057, assuming $M_{NS}=1.4\:M_{\odot}$ and $R_{NS}=10$ km. The extent of the boundary layer region for both sources is perfectly consistent with the position of the inner accretion disc. However, the actual value may be larger than this if we account for the spin and the viscous effects in the layer. \\

The inferred inner disc radius suggests the disc is probably truncated moderately away from the NS surface. It has been observed that
for X-ray pulsars, the magnetic field of an NS can potentially truncate the inner disc and re-direct plasma along the magnetic field lines. We can estimate the magnetic field strength if it is truncated at magnetospheric radius. The upper limit of $R_{in}$ measured from the reflection fit can be used to estimate an upper limit of the magnetic field strength of the NS. We have used the following expression for the calculation of the magnetic dipole moment \citep{2009ApJ...694L..21C},
\begin{equation}
\begin{split}
\mu=&3.5\times 10^{23}k_{A}^{-7/4} x^{7/4} \left(\frac{M}{1.4 M_{\odot}}\right)^{2}\\
 &\times\left(\frac{f_{ang}}{\eta}\frac{F_{bol}}{10^{-9} \text{erg}\: \text{cm}^{-2}\: \text{s}^{-1}}\right)^{1/2}
 \frac{D}{3.5\: \text{kpc}} \text{G}\; \text{cm}^{3}.
\end{split} 
\end{equation}
Here $k_{A}$ is the geometrical coefficient, $f_{ang}$ is the anisotropy correction factor, and $\eta$ is the accretion efficiency in the Schwarzschild metric. We assumed $k_{A}=1$, $f_{ang}=1$, and $\eta=0.2$ (as reported in \citealt{2009ApJ...694L..21C} and \citealt{2000AstL...26..699S}). We used $0.1 - 100$\kev{} flux as the bolometric flux ($F_{bol}$) of $\sim 8.84\times 10^{-9}$ erg~s$^{-1}$ cm$^{-2}$ for the source Swift~J17480 and $\sim 7.83\times 10^{-10}$ erg~s$^{-1}$ cm$^{-2}$ for the source IGR~J17511-3057. Here, $R_{in}$ is modified as $R_{in}=x\:GM/c^{2}$ by introducing a scale factor $x$ (\citet{2009ApJ...694L..21C}). Using the upper limit of $R_{in} \lesssim 16\:R_{g}$ for the source Swift~J17480 and $\lesssim 7.3\:R_{g}$ for the source IGR~J17511-3057, we obtained $\mu \leq 0.5\times 10^{27}$ G cm$^{3}$ and $\leq 0.05\times 10^{27}$ G cm$^{3}$ for the source Swift~J17480 and IGR~J17511-3057, respectively. These leads to an upper limit of the magnetic field strength of $B\lesssim 1.0\times 10^{9}$ G and $B\lesssim 0.96\times 10^{8}$ G at the magnetic poles for the source Swift~J17480 and IGR~J17511-3057, respectively (assuming an NS mass of $1.4\:M_{\odot}$, a radius of $10$ km, and a distance of $5.9$ kpc for the source Swift~J17480 and $6.9$ kpc for the source IGR~J17511-3057). The magnetic field strength of Swift~J17480 is consistent with the previously estimated magnetic field strength of other Terzan 5 source IGR~J17480-2446 (\citealt{2011ApJ...731L...7M, 2011A&A...526L...3P}). The magnetic field strength of IGR~J17511-3057 is perfectly consistent with the estimation of \citet{2015MNRAS.452.3994M}. All the physical parameters inferred from the spectral fitting are listed in Table~\ref{table:4}.\\

\section{Data availability}
This research has made use of data obtained from the HEASARC, provided by NASA's Goddard Space Flight Center. The observational data sets with Obs. IDs $80601304002$ (\nustar{}) dated March 4, 2023 and $90101001002$ (\nustar{}) dated April 8, 2015 are in public domain put by NASA at their website https://heasarc.gsfc.nasa.gov.  
 
\section{Acknowledgements}
We thank the referee for his/her expert comments/suggestions that improved the clarity of the presentation and description of the paper. We express our sincere gratitude to Craig Heinke for the explanation related to the active transients in the globular cluster Terzan 5. We also thank him for sharing the plots included in Figure~1 during a private communication. This research has made use of data and/or software provided by the High Energy Astrophysics Science Archive Research Centre (HEASARC). This research also has made use of the \nustar{} data analysis software ({\tt NuSTARDAS}) jointly developed by the ASI Space Science Data Center (SSDC, Italy) and the California Institute of Technology (Caltech, USA). ASM and BR would like to thank Inter-University Centre for Astronomy and Astrophysics (IUCAA) for their facilities extended to him under their Visiting Associate Programme.

\def\apj{ApJ}
\def\apjl{ApJl}
\def\pasp{PASP} \def\mnras{MNRAS} \def\aap{A\&A} \def\physerp{PhR} \def\apjs{ApJS} \def\pasa{PASA}
\def\pasj{PASJ} \def\nat{Nature} \def\memsai{MmSAI} \def\araa{ARAA} \def\iaucirc{IAUC} \def\aj{AJ} \def\aaps{A\&AS} \def\apss{APSS} \def\nar{NewAR}
\bibliographystyle{unsrt}
\bibliography{aditya}

\end{document}